\documentclass[12pt, onecolumn]{IEEEtran}

\usepackage{setspace}
\usepackage[a4paper]{geometry}
\usepackage{amsmath}
\usepackage{amsthm}
\usepackage{amssymb}
\usepackage{verbatim}
\usepackage{algorithm}
\usepackage{algorithmic}
\usepackage{graphicx}
\usepackage{paralist}
\usepackage{color}

\theoremstyle{plain}

\newtheorem{theorem}{Theorem}
\newtheorem{definition}{Definition}

\theoremstyle{definition}
\newtheorem{counterexample}{Counterexample}
\floatname{algorithm}{Algorithm}
\newtheorem*{remark}{Remark}

\allowdisplaybreaks[4]

\begin{document}
%
\title{On the Optimality of Myopic Sensing \\ in Multi-channel Opportunistic Access: \\ the Case of Sensing Multiple Channels}
%
%
%
%

\author{Kehao~Wang, \qquad Lin~Chen
\IEEEcompsocitemizethanks{\IEEEcompsocthanksitem The authors are with the Laboratoire de Recherche en Informatique (LRI), Department of Computer Science, the University of Paris-Sud XI, 91405 Orsay, France (e-mail: \{Kehao.Wang, Lin.Chen\}@lri.fr).}
}

\addtolength{\abovedisplayskip}{-1ex}
\addtolength{\belowdisplayskip}{-1ex}

\IEEEcompsoctitleabstractindextext{%
\begin{abstract}
Recent works (\cite{Qzhao08,Sahmand09}) have developed a simple and robust myopic sensing policy for multi-channel opportunistic communication systems where a secondary user (SU) can access one of $N$ i.i.d. Markovian channels. The optimality of the myopic sensing policy in maximizing the SU's cumulated reward is established under certain conditions on channel parameters. This paper studies the generic case where the SU can sense more than one channel each time. By characterizing the myopic sensing policy in this context, we establish analytically its optimality for certain system setting when the SU is allowed to sense two channels. In the more generic case, we construct counterexamples to show that the myopic sensing policy, despite its simple structure, is non-optimal.
\end{abstract}

}

\maketitle

\IEEEdisplaynotcompsoctitleabstractindextext

%
\IEEEpeerreviewmaketitle

\section{Introduction}

The concept of opportunistic spectrum access (OSA), first envisioned by J. Mitola in the seminal paper \cite{Mitola99} on the software defined radio systems, has emerged in recent years as a
promising paradigm to enable more efficient spectrum utilization. The basic idea of OSA is to exploit instantaneous spectrum availability by allowing the unlicensed secondary users (SU) to access the temporarily unused channels of the licensed primary users (PU) in an opportunistic fashion. In this context, a well-designed channel access policy is crucial to achieve efficient spectrum usage.

\begin{figure}[htbp]
\centering
\includegraphics[width=8cm]{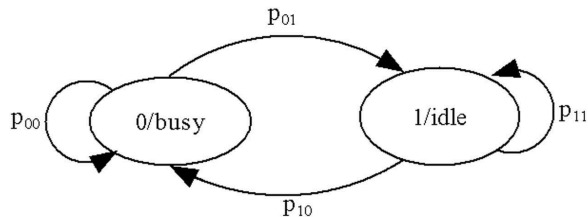}
\caption{The Markov channel model}
\label{fig:channel_model}
\end{figure}

In this paper, we consider a generic OSA scenario where there are $N$ slotted spectrum channels partially occupied by the PUs. Each channel evolves as an independent and identically distributed (i.i.d.), two-state discrete-time Markov chain. As illustrated in Fig.~\ref{fig:channel_model}, the two states for each channel, busy (state $0$) and idle (state $1$), indicate whether the channel is free for an SU to transmit its packet on that channel at a given slot. The state transition probabilities are given by $\{p_{ij}\}, i,j=0,1$. An SU seeks a sensing policy that opportunistically exploits the temporarily unused channels to transmit its packets. To this end, in each slot, the SU selects a subset of channels to sense based on its prior observations and obtain one unit as reward if at least one of the sensed channel is in the idle state, indicating that the SU can effectively send one packet using the idle channel (or one of the idle channels) unused by PUs in current slot. The objective of the SU is to find the optimal sensing policy maximizing the reward that it can obtain over a finite or infinite time horizon.

As stated in~\cite{Qzhao08}, the design of the optimal sensing policy can be formulated as a partially observable Markov decision
process (POMDP)~\cite{Zhao07}, or a restless multi-armed bandit problem (RMAB)~\cite{Whittle88}, of which the application is far beyond the domain of cognitive radio systems\footnote{Please refer to~\cite{Qzhao08,Sahmand09} for more examples where this formulation is applicable. A summary on the related works on the analysis of this problem using the POMDP and RMAB approaches are presented in~\cite{Sahmand09}. We thus do not provide a literature survey in this paper.}. Unfortunately, obtaining the optimal policy for a general POMDP or RMAB is often intractable due to the exponential computation complexity. Hence, a natural alternative is to seek simple myopic policies for the SU. In this line of research, a myopic sensing strategy is developed in~\cite{Qzhao08} for the case where the SU is limited to sense only one channel at each slot. The myopic sensing policy in this case is proven to be optimal when the state transitions of the Markov channels are positively correlated, i.e., $p_{11}\ge p_{01}$~\cite{Sahmand09}.

In this paper, we naturally extend the proposed myopic policy in the generic case where the SU can sense more than one channel each time and gets one unit of reward if at least one of the sensed channels is in the idle state. Through  mathematic analysis, we show that the generalized myopic sensing policy is optimal only for a small subset of cases where the SU is allowed to sense two channels each slot. In the general case, we give counterexamples to show that the myopic sensing policy, despite its simple structure, is not optimal.
We believe that our results presented in this paper, together with~\cite{Qzhao08,Sahmand09,Ahmad09}, lead to more in-depth understanding of the intrinsic structure and the resulting optimality of the myopic sensing policy and will stimulate more profound research on this topic.

Before concluding the Introduction section, it is insightful to compare our results obtained in this paper with that presented in~\cite{Ahmad09} on the similar problem. In~\cite{Ahmad09}, the authors show that when $p_{11}\ge p_{01}$ holds, the myopic sensing policy is optimal even for the case where the SU senses more than one channel each slot. These two results seem to be contradictory. In fact, this is due to the fact that in~\cite{Ahmad09}, the objective of the SU is to find as many idle channels as possible, thus maximizing the throughput under the condition that the SU can transmit on all the idle channels. In contrast, our results are focused on the scenario where the SU can only transmit on one channel. As a result, the SU aims at maximizing the probability to find at least one idle channel. It is insightful to notice that this nuance on the model (more specifically on the utility function) indeed leads to totally contrary results, indicating that more research efforts are needed to understand the intrinsic characteristics of the myopic policy. In fact, we are currently investigating the forms of utility function under which the myopic policy can be optimal.

The rest of this paper is structured as follows. Section~\ref{sec:problem_formulation} formulates the optimization channel sensing problem for the SU and presents the myopic sensing policy in the generic case. Section~\ref{sec:myopic_sensing} studies the optimality of the myopic sensing policy, with Subsection~\ref{subsec:optimality_myopic_sensing} establishing the optimality of the myopic sensing policy for a subset of scenarios when the SU is allowed to sense two channels each time, and Subsection~\ref{subsec:non_optimality_myopic_sensing} illustrating the non-optimality of the myopic sensing policy for the general case through two representative counterexamples. Finally, the paper is concluded in Section~\ref{sec:conclusion}.

\section{Problem Formulation}
\label{sec:problem_formulation}

As explained in the Introduction, we are interested in a synchronously slotted cognitive radio network where an SU can opportunistically access a set $\mathcal{N}$ of $N$ i.i.d. channels partially occupied by PUs. The state of each channel $i$ in time slot $t$, denoted by $S_{i}(t)$, is modeled by a discrete time two-state Markov chain shown in Fig 1. At the beginning of each slot $t$, the SU selects a subset $\mathcal{A}(t)$ of channels to sense. If at least one of the sensed channels is in the idle state (i.e., unoccupied by any PU), the SU transmits its packet and collects one unit of reward. Otherwise, the SU cannot transmit, thus obtaining no reward. These decision procedure is repeated for each slot. The focus of our work is to study the optimal sensing policy of the SU in order to maximize the average reward over $T$ slots. Let $\mathbf{{\mathcal A}}=\{{\mathcal A}(t), 1\le t\le T\}$, the optimization problem of the SU $P_{SU}$, when the SU is allowed to sense $k$ channels, is formally defined as follows\footnote{The more generic utility function can be formed by integrating the discount factor and allowing $T=+\infty$. By slightly adapting the analysis in this paper, our analysis can be extended there.}:
\begin{equation}
P_{SU}: \ \max_{{\mathcal A}(t)\subseteq\mathcal{N}, \ |{\mathcal A}(t)|=k, \ 0\le t\le T-1} \quad \frac{1}{T}\sum_{t=1}^T\left[1-\prod_{i\in{\mathcal A}(t)}(1-\omega_i(t))\right],
\end{equation}
where $\omega_i(t)$ is the conditional probability that $S_i(t)=1$ given the past actions and observations\footnote{$\omega_i(0)$ can be set to $\frac{p_{01}}{p_{01}+p_{11}}$ if no information about the initial system state is available.}. Based on the sensing policy $\mathcal{A}(t)$ in slot $t$ and the sensing result, $\{\omega_i(t), i\in{\cal N}\}$ can be updated using Bayes Rule as shown in~\eqref{eq:belief_update}.
\begin{equation}
\omega_i(t+1)=
\begin{cases}
p_{11}, & i\in\mathcal{A}(t), S_i(t)=1 \\
p_{01}, & i\in\mathcal{A}(t), S_i(t)=0 \\
\tau(\omega_i(t)), & i\not\in\mathcal{A}(t)
\end{cases},
\label{eq:belief_update}
\end{equation}
where $\tau(\omega_i(t))\triangleq\omega_i(t)p_{11}+[1-\omega_i(t)]p_{01}$ characterizes the evolution of the believe value of the non-sensed channels.

As argued in the Introduction, the optimization problem $P_{SU}$ is by nature a POMDP, or a restless multi-armed bandit problem, of which the optimal sensing policy is in general intractable. Hence, a natural alternative is to seek simple myopic sensing policy, i.e., the sensing policy maximizing the immediate reward based on current believe. In this line of research, a myopic sensing policy is developed in~\cite{Qzhao08} for the case where the SU is limited to sense only one channel at each slot. The goal of our work presented in this paper is to study the optimality of the myopic sensing policy in the generic case where the SU can sense multiple channels each time. To this end, we first derive the structure of the myopic sensing policy for the general case and then provide an in-depth analysis on its optimality in Section~\ref{sec:myopic_sensing}.


\begin{definition}[Structure of Myopic Sensing in Generic Case]
Sort the elements of the belief vector in descending order such that $\omega_1(t)\ge \omega_2(t)\ge \cdots \ge \omega_N(t)$, the myopic sensing policy in the generic case, where the SU is allowed to sense $k$ channels, consists of sensing channel $1$ to channel $k$.
\end{definition}

The myopic sensing policy is easy to implement and maximizes the immediate payoff. In the next section, we show that the myopic sensing policy is optimal for the case $k=2$ and $T=2$ when $p_{11}\ge p_{01}$ and when $p_{11}<p_{01}$ and $N\le 4$. Beyond this small subset of parameter settings, we show that the myopic sensing policy, despite its simple structure, is not optimal by constructing two representative counterexamples. 

\section{Optimality of Myopic Sensing Policy}
\label{sec:myopic_sensing}

In this section, we study the optimality of the myopic sensing policy for the generic case ($k\ge 2$). More specifically, we structure our analysis into two cases: (1) $T=2$, $k=2$; (2) the general case.

\subsection{Optimality of myopic sensing policy when $T=2$ and $k=2$}
\label{subsec:optimality_myopic_sensing}

This subsection is focused on the case where the SU is allowed to sense two channels each slot and aims at maximizing the reward of the upcoming two slots. In terms of user behavior, this case models a short-sighted SU. The following two theorems study the optimality of the myopic sensing policy in this case for $p_{11}\ge p_{01}$ and $p_{11}<p_{01}$, respectively.

\begin{theorem}[Optimality of Myopic Sensing Policy for $T=2$ and $k=2$: $p_{11}\ge p_{01}$]
In the case where $T=2$ and $k=2$, the myopic sensing policy is optimal when $p_{11}\ge p_{01}$.
\label{theorem:optimality_myopic_positive}
\end{theorem}

\begin{proof}
We sort the elements of the believe vector at the beginning of the slot $t$ $[\omega_1(t), \omega_2(t), \cdots, \omega_N(t)]$ in descending order such that $\omega_1\ge\omega_2\ge\cdots\ge\omega_N$\footnote{For the simplicity of presentation, by slightly abusing the notations without introducing ambiguity, we drop the time slot index of $\omega_i(t)$.}.
Under this notation, we can write the reward of the myopic sensing policy (i.e., sensing channel $1$ and $2$), denoted as $R^*$, as
\begin{multline}
R^*= \underbrace{1-(1-\omega_1)(1-\omega_2)}_{A} + \underbrace{\omega_1\omega_2[1-(1-p_{11})(1-p_{11})]}_{B} + \underbrace{\omega_1(1-\omega_2)[1-(1-p_{11})(1-\tau(\omega_3))]}_{C} + \\ \underbrace{(1-\omega_1)\omega_2[1-(1-p_{11})(1-\tau(\omega_3))]}_{D} + \underbrace{(1-\omega_1)(1-\omega_2)[1-(1-\tau(\omega_3)(1-F))]}_{E},
\label{eq:reward_myopic}
\end{multline}
where $\tau(\omega)=\omega p_{11}+(1-\omega)p_{01}$ is defined in~\eqref{eq:belief_update}, $F=p_{01}$ when $N=3$ and $\tau(\omega_4)$ when $N\ge 4$. More specifically, term $A$ denotes the immediate reward in the upcoming slot $t$; term $B$ denotes the expected reward of slot $t+1$ when both channels are sensed to be free; term $C$ (term $D$, respectively) denote the expected reward of slot $t+1$ when only channel $1$ (channel 2) is sensed to be free; term $E$ denotes the expected reward of slot $t+1$ when both channels are sensed to be busy.

The proof of Theorem~\ref{theorem:optimality_myopic_positive} consists of showing that sensing any two channels $\{i,j\}\ne\{1,2\}$ cannot bring the SU more reward. We proceed the proof for the following two cases:
\begin{itemize}
\item $\{i,j\}$ is partially overlapped with $\{1,2\}$, i.e., $\{i,j\}\bigcap\{1,2\}\ne\emptyset$;
\item $\{i,j\}$ is totally distinct to $\{1,2\}$, i.e., $\{i,j\}\bigcap\{1,2\}=\emptyset$;
\end{itemize}

\textbf{Case 1.} When $\{i,j\}$ is partially overlapped with $\{1,2\}$, without loss of generality, assume that $i=1$ and $j\ge 3$, we can derive the upper bound of the expected reward of sensing the channels $\{i,j\}=\{1,j\}$, as shown in equation \eqref{eq:reward_case1j3} (when $j=3$) and equation \eqref{eq:reward_case1j4} (when $j\ge4$). Here by upper bound we mean that the SU senses channel $i$ and $j$ in slot $t$ and the two channels with the largest idle probabilities for slot $t+1$, leading to the maximal reward that the SU can achieve.

When $j=3$, following the similar analysis as that in \eqref{eq:reward_myopic}, the utility upper bound when sensing the channels $\{i,j\}=\{1,3\}$, denoted by $\overline{R}_1$, can be derived as follows:
\begin{multline}
\overline{R}_1= 1-(1-\omega_1)(1-\omega_j) + \omega_1\omega_j[1-(1-p_{11})(1-p_{11})] + \omega_1(1-\omega_j)[1-(1-p_{11})(1-\tau(\omega_2))] + \\ (1-\omega_1)\omega_j[1-(1-p_{11})(1-\tau(\omega_2))] + (1-\omega_1)(1-\omega_j)[1-(1-\tau(\omega_2)(1-p_{01}))].
\label{eq:reward_case1j3}
\end{multline}

After some algebraic operations, we obtain
\begin{equation}
R^*-R_1 = (1-\omega_1)(\omega_2-\omega_3)(1-(1-p_{11})(F-p_{01})),
\label{eq:jeq3}
\end{equation}
where $F$ is defined in~\eqref{eq:reward_myopic}. Noticing that $p_{01}\le\tau(\omega_i)\le p_{11},\forall i\in{\cal N}$ following \eqref{eq:belief_update}, it holds that $F\ge p_{01}$. Hence $R^*-\overline{R}_1\ge 0$ holds for $j=3$.

When $j\ge4$, the utility upper bound can be derived as:
\begin{multline}
\overline{R}_1= 1-(1-\omega_1)(1-\omega_j) + \omega_1\omega_j[1-(1-p_{11})(1-p_{11})] + \omega_1(1-\omega_j)[1-(1-p_{11})(1-\tau(\omega_2))] + \\ (1-\omega_1)\omega_j[1-(1-p_{11})(1-\tau(\omega_2))] + (1-\omega_1)(1-\omega_j)[1-(1-\tau(\omega_2)(1-\tau(\omega_3)))].
\label{eq:reward_case1j4}
\end{multline}

It follows that
\begin{multline}
\label{eq:jgeq4}
R^*-\overline{R}_1= \omega_1(1-\omega_2)(\omega_3-\omega_j)(p_{11}-p_{01}) + (1-\omega_1)(\tau(\omega_2)-\tau(\omega_j))(\omega_2(1-p_{11})+ \\ (1-\tau(\omega_3))(1-\omega_2))+
(1-\omega_1)(\tau(\omega_2)-\tau(\omega_j))(1-(1-p_{11})(\tau(\omega_3)-p_{01})),
\end{multline}

It follows from the definition of $\tau(\omega)$ after~\eqref{eq:belief_update} that when $p_{01}\le p_{11}$, $\tau(\omega)$ is increasing in $\omega$ and that $p_{01}\le\tau(\omega_3)\le p_{11}$. Hence $R^*-\overline{R}_1\ge 0$ holds for $j\ge 4$, too.

The above results show that any other sensing policy cannot outperform the myopic sensing policy in this case.

\textbf{Case 2.} When $\{i,j\}$ is totally distinct to $\{1,2\}$, implying $N\ge 4$, we can write the reward upper bound of the sensing policy $\{i,j\}$, denoted as $\overline{R}_2$, in \eqref{eq:reward_case2}:
\begin{multline}
\overline{R}_2= 1-(1-\omega_i)(1-\omega_j) + \omega_i\omega_j[1-(1-p_{11})(1-p_{11})] + \omega_i(1-\omega_j)[1-(1-p_{11})(1-\tau(\omega_1))] + \\ (1-\omega_j)\omega_j[1-(1-p_{11})(1-\tau(\omega_1))] + (1-\omega_i)(1-\omega_j)[1-(1-\tau(\omega_1)(1-\tau(\omega_2))].
\label{eq:reward_case2}
\end{multline}

We take $\overline{R}_1$ in \eqref{eq:reward_case1j4} as an auxiliary to proceed our mathematic analysis. More specifically, comparing $\overline{R}_2$ to $\overline{R}_1$ in \eqref{eq:reward_case1j4}, after some algebraic operations, we have
\begin{multline}
\overline{R}_1-\overline{R}_2= (1-\omega_j)(\omega_1-\omega_i)+  \omega_j(\omega_1-\omega_i)(p_{11}+p_{11}-(p_{11})^2) + \\
          (1-\omega_j)[p_{11}(\omega_1-\omega_i)+(1-p_{11})(\omega_1\tau(\omega_2)-\tau(\omega_1)\omega_i)] +\\
              \omega_j[(1-p_{11})((1-\omega_1)\tau(\omega_2)-(1-\omega_i)\tau(\omega_1))-p_{11}(\omega_1-\omega_i)]+\\
              (1-\omega_j)[-(\omega_1-\omega_i)+(1-\tau(\omega_2))[(1-\tau(\omega_1))(1-\omega_i)-(1-\omega_1)(\tau(\omega_3))]]\\
              \geq (1-\omega_j)(\omega_1-\omega_i)+  \omega_j(\omega_1-\omega_i)(p_{11}+p_{11}-(p_{11})^2) + \\
             (1-\omega_j)[p_{11}(\omega_1-\omega_i)+(1-p_{11})(\omega_1\tau(\omega_i)-\tau(\omega_1)\omega_i)] +\\
              \omega_j[(1-p_{11})((1-\omega_1)\tau(\omega_i)-(1-\omega_i)\tau(\omega_1))-p_{11}(\omega_1-\omega_i)]+\\
              (1-\omega_j)[-(\omega_1-\omega_i)+(1-\tau(\omega_2))[(1-\tau(\omega_1))(1-\omega_i)-(1-\omega_1)(\tau(\omega_i))]]\\
              = (1-\omega_j)(\omega_1-\omega_i)[p_{11}+(1-p_{11})p_{01}+(1-p_{11})(1-\tau(\omega_2))] \geq 0.
\end{multline}

It then follows that
\begin{equation}
R^*-\overline{R}_2= (R^*-\overline{R}_1) + (\overline{R}_1-\overline{R}_2) \geq 0,
\end{equation}
meaning that sensing $\{i,j\}$ cannot outperform the myopic sensing policy in this case, either. Combining the results of both cases completes the proof of Theorem~\ref{theorem:optimality_myopic_positive}.
\end{proof}

The following theorem studies the optimality of the myopic sensing policy when $p_{11}<p_{01}$. The proof follows the similar way as that of Theorem~\ref{theorem:optimality_myopic_positive} and is thus omitted.

\begin{theorem}[Optimality of Myopic Sensing Policy for $T=2$ and $k=2$: $p_{11}<p_{01}$]
In the case where $T=2$ and $k=2$, the myopic sensing policy is optimal when $p_{11}<p_{01}$ for the system consisting of at most $4$ channels (i.e., $N\le 4$).
\label{theorem:optimality_myopic_negative}
\end{theorem}

The optimality of the myopic sensing policy derived in this subsection, especially when $p_{11}\ge p_{01}$, hinges on the fact that the eventual loss of reward in slot $t+1$, if there is, is over compensated by the reward gain in the current slot $t$. However, this result cannot be iterated in the general case. On the contrary, in the next subsection, we show that the myopic sensing policy may not be optimal by providing two representative counterexamples.

\subsection{Non-optimality of myopic sensing policy in general cases}
\label{subsec:non_optimality_myopic_sensing}

In this subsection, we show that the myopic sensing policy is not optimal for the general cases beyond those studied in Section~\ref{subsec:optimality_myopic_sensing} by constructing two representative counterexamples.

\begin{counterexample}[$k=3$, $T=2$, $N=6$, $p_{11}\ge p_{01}$]
Consider a system with $k=3$, $T=2$, $N=6$ and $p_{11}>p_{01}$, the reward generated by the myopic sensing policy (sensing the $3$ channels with highest elements in the believe vector at each slot, i.e., $\omega_1$, $\omega_2$, $\omega_3$) is given by
\begin{multline}
R^*_{c1}= 1-(1-\omega_1)(1-\omega_2)(1-\omega_3) + \omega_1\omega_2\omega_3[1-(1-p_{11})^3] + \\ [\omega_1\omega_2(1-\omega_3)+\omega_1(1-\omega_2)\omega_3+(1-\omega_1)\omega_2\omega_3][1-(1-p_{11})^2(1-\tau(\omega_4))] + \\ [\omega_1(1-\omega_2)(1-\omega_3)+(1-\omega_1)\omega_2(1-\omega_3)+(1-\omega_1)(1-\omega_2)\omega_3][1-(1-p_{11})(1-\tau(\omega_4))(1-\tau(\omega_5))] + \\ (1-\omega_1)(1-\omega_2)(1-\omega_3)[1-(1-\tau(\omega_4)(1-\tau(\omega_5))(1-\tau(\omega_6))].
\end{multline}
On the other hand, consider the sensing policy that senses the $2$ highest elements and the forth highest element in the believe vector (i.e., $\omega_1$, $\omega_2$ and $\omega_4$ according ot our notation) for the current slot $t$ and senses the highest $3$ elements in the believe vector for slot $t+1$, the reward generated by this policy is
\begin{multline}
R_{c1}= 1-(1-\omega_1)(1-\omega_2)(1-\omega_4) + \omega_1\omega_2\omega_4[1-(1-p_{11})^3] + \\ [\omega_1\omega_2(1-\omega_4)+\omega_1(1-\omega_2)\omega_4+(1-\omega_1)\omega_2\omega_4][1-(1-p_{11})^2(1-\tau(\omega_3))] + \\ [\omega_1(1-\omega_2)(1-\omega_4)+(1-\omega_1)\omega_2(1-\omega_4)+(1-\omega_1)(1-\omega_2)\omega_4][1-(1-p_{11})(1-\tau(\omega_3))(1-\tau(\omega_5))] + \\ (1-\omega_1)(1-\omega_2)(1-\omega_4)[1-(1-\tau(\omega_3)(1-\tau(\omega_5))(1-\tau(\omega_6))].
\end{multline}
It can be calculated that under the setting $[\omega_1, \omega_2, \omega_3, \omega_4, \omega_5, \omega_6]=[0.99, 0.5, 0.4, 0.39, 0.25, 0.25]$, $p_{11}=0.5$, $p_{01}=0.3$, it holds that $R_{c1}-R^*_{c1}=0.00005625>0$. The myopic sensing policy is not optimal for this counterexample.
\end{counterexample}

\begin{counterexample}[$k=3$, $T=2$, $N=6$, $p_{11}<p_{01}$]
In the case of $p_{11}<p_{01}$, $k=3$, $T=2$ and $N=6$, the reward generated by the myopic sensing policy is:
\begin{multline}
R^*_{c2}= 1-(1-\omega_1)(1-\omega_2)(1-\omega_3) + \omega_1\omega_2\omega_3[1-(1-\tau(\omega_6))(1-\tau(\omega_5))(1-\tau(\omega_4))] + \\ [\omega_1\omega_2(1-\omega_3)+\omega_1(1-\omega_2)\omega_3+(1-\omega_1)\omega_2\omega_3][1-(1-p_{01})(1-\tau(\omega_6))(1-\tau(\omega_5))] + \\ [\omega_1(1-\omega_2)(1-\omega_3)+(1-\omega_1)\omega_2(1-\omega_3)+(1-\omega_1)(1-\omega_2)\omega_3][1-(1-p_{01})^2(1-\tau(\omega_6))] + \\ (1-\omega_1)(1-\omega_2)(1-\omega_3)[1-(1-p_{01})^3].
\end{multline}

The reward generated by the strategy of sensing channels $1,2,4$ is:
\begin{multline}
R_{c2}= 1-(1-\omega_1)(1-\omega_2)(1-\omega_4) + \omega_1\omega_2\omega_4[1-(1-\tau(\omega_6))(1-\tau(\omega_5))(1-\tau(\omega_3))] + \\ [\omega_1\omega_2(1-\omega_4)+\omega_1(1-\omega_2)\omega_4+(1-\omega_1)\omega_2\omega_4][1-(1-p_{01})(1-\tau(\omega_6))(1-\tau(\omega_5))] + \\ [\omega_1(1-\omega_2)(1-\omega_4)+(1-\omega_1)\omega_2(1-\omega_4)+(1-\omega_1)(1-\omega_2)\omega_4][1-(1-p_{01})^2(1-\tau(\omega_6))] + \\ (1-\omega_1)(1-\omega_2)(1-\omega_3)[1-(1-p_{01})^3].
\end{multline}
We have $R_{c2}-R^*_{c2}=0.00002>0$ with the parameters $[\omega_1, \omega_2, \omega_3, \omega_4, \omega_5, \omega_6]=[0.99, 0.5, 0.4, 0.39, 0.25, 0.25]$, $p_{11}=0.3$, $p_{01}=0.5$. The myopic sensing policy is not optimal for this counterexample, either.
\end{counterexample}

\begin{remark}
The above counterexamples can serve as a basis to construct more general counterexamples for the case where $T\ge 3$. One such counterexample in the general case where $T\geq 3$ is to follow the sensing policy given in the counterexample that gives better two-slot reward and than follow the optimal sensing policy. As a result, the global reward of the constructed sensing policy outweighs that of the myopic sensing policy.
\end{remark}

We are now ready to state the major result in this paper.

\begin{theorem}[Non-optimality of Myopic Sensing Policy in General Case]
The myopic sensing policy is not guaranteed to be optimal in the general case.
\label{theorem:non_optimality_myopic}
\end{theorem}

\begin{remark}
To conclude this section, it is insightful to note that the major results of this paper on the optimality of the myopic sensing policy, stated in Theorem~\ref{theorem:optimality_myopic_positive}, Theorem~\ref{theorem:optimality_myopic_negative} and Theorem~\ref{theorem:non_optimality_myopic}, hinge on the fundamental trade-off between exploration, by sensing unexplored channels in order to learn and predict the future channel state, thus maximizing the long-term reward (e.g., term $B,C,D,E$ in~\eqref{eq:reward_myopic}), and exploitation, by accessing the channel with the highest estimated idle probability based on currently available information (the belief vector) which greedily maximizes the immediate reward (e.g., term $A$ in~\eqref{eq:reward_myopic}). For a short-sighted SU ($T=1$ and $T=2$), exploitation naturally dominates exploration (i.e., the immediate reward overweighs the potential gain in future reward) under certain system parameter settings, resulting the optimality of the myopic sensing policy in a subset of this scenario. In contrast, to achieve maximal reward for $T\ge 3$, the SU should strike a balance between exploration and exploitation. In such context, the myopic sensing policy that greedily maximizes the immediate reward is no more optimal.
\end{remark}

\section{Conclusion}
\label{sec:conclusion}

In this paper, we study the optimality of the myopic sensing policy in the generic scenario of opportunistic spectrum access in a multi-channel communication system where an SU senses a subset of channels partially occupied by licensed PUs. We show that the myopic sensing policy is optimal only for a small subset of cases where the SU is allowed to sense two channels each slot. In the generic case, we give counterexamples to show that the myopic sensing policy, despite its simple structure, is not optimal. Due to the generic nature of the problem, we believe that the results obtained in this paper leads to more in-depth understanding of the intrinsic structure and the resulting optimality of the myopic policy and will stimulate more profound research on this topic.

\bibliographystyle{unsrt}
\bibliography{reference}

\end{document}